# Spin-group theory on Edelstein effect and spin-orbit torque in Collinear Ferromagnets

Yizhuo Song[1], Qing Zhang[1], Jiahao Shentu[1], Jie Li[1], and Jia Zhang[1*]

[1]*School of Physics and Wuhan National High Magnetic Field Center, Huazhong University of Science and Technology, 430074 Wuhan, China*

[*]jiazhang@hust.edu.cn.

## Abstract

Current-induced spin-orbit torques (SOTs) are central to the electrical manipulation of magnetic order in spintronic devices. In transition-metal/collinear ferromagnet bilayers, field-like and damping-like torques have been described only phenomenologically via the spin or orbital Hall effect, lacking a rigorous symmetry-based foundation. The precise role of spin-orbit coupling (SOC) in both the Edelstein effect and SOTs has remained unresolved. Here we develop a spin-group symmetry theory for the Edelstein effect and SOTs in collinear ferromagnets, treating SOC as a symmetry-breaking perturbation. For $4mm$ ($C_{4v}$) point group symmetry, we derive the full forms of field-like and damping-like torques, which arise predominantly from first- and second-order SOC. We further show that SOTs in both orbital-Hall-dominated Ti/Ni and spin-Hall-dominated Pt/CoFe bilayers originate at first-order SOC. Taking the $3m$ ($C_{3v}$) torque as a paradigmatic example, we elucidate the role of second- and higher-order SOC torques in field-free switching of perpendicular magnetic anisotropy. Remarkably, in PtMnSb, we demonstrate that SOTs under certain point group symmetries deviate from the conventional form: zeroth- and first-order SOC contributions vanish identically, with the leading SOT emerging at second order. All symmetry-based predictions from spin-group theory are in excellent quantitative agreement with first-principles calculations. Our work establishes a unified symmetry framework for the microscopic understanding of the Edelstein effect and current-induced spin torques in ferromagnetic systems.

## I. Introduction

Current-induced spin-orbit torque (SOT) enables all-electrical control of magnetic order and underpins next-generation spintronic devices such as magnetic random-access memory (MRAM)[1] and spin logic circuits. Decades of intensive studies have sought to unravel the microscopic origins of SOT, firmly establishing spin-orbit coupling (SOC) as the primary driving force behind spin-charge interconversion. At the heart of the interconversion processes lies the Edelstein effect[2], a paradigmatic linear-response phenomenon in which an applied electric field induces a non-equilibrium spin density in systems with broken inversion symmetry. Through exchange interaction to the local magnetic moments, this field-driven spin accumulation exerts a torque on the magnetization, forming the microscopic basis of SOT.

Conventional theoretical treatments of the Edelstein effect and the resulting SOT typically rely on an expansion in the magnetic order parameter $\boldsymbol{m}$[3], yet the relative magnitudes of distinct contributions and their scaling with SOC strength remain unresolved. In particular, in transition-metal/ferromagnet bilayers, SOT may arise from either the spin or orbital Hall effect in the transition-metal layer[1][4][5][6]. However, it remains unclear whether these two distinct mechanisms share a common scaling behavior with SOC, and to what extent a low-order SOC expansion remains quantitatively valid. Resolving this issue is critical not only for accurate theoretical modeling of SOT phenomena but also for the rational design of spintronic materials with tailored torque efficiencies.

In this work, we develop a spin-group symmetry theory of the Edelstein effect and the resulting SOT in collinear ferromagnetic systems formulated with spin-orbit vectors. We focus on crystallographic point groups of broad practical relevance, particularly the 4*mm* ($C_{4v}$) and 3*m* ($C_{3v}$) point group, which describes a wide class of ferromagnetic bilayer heterostructures and also applies to polycrystalline films with $C_{\infty v}$ symmetry. By enforcing spin-group symmetry constraints, we derive the explicit forms of the electric field-induced spin density as well as the field-like and damping-like torkances, establishing a clear mapping between the SOC perturbation order and the corresponding response.

We validate our symmetry theory via first-principles calculations on three prototypical bilayer systems including $C_{4v}$ system Ti/Ni(001) bilayers dominated by the orbital Hall effect, and Pt/CoFe(001) bilayers governed by the spin Hall effect, as well as $C_{3v}$ system Pt(111)/Co(111). We further verify the framework in a cubic bulk ferromagnet PtMnSb with vanishing first-order SOC Edelstein effect and absence of conventional SOT. Our work establishes a unified framework for classifying SOT responses by SOC order, and offers practical guidance for the theoretical modeling and experimental interpretation of SOT across diverse ferromagnetic systems.

## II. SOC expansion of Edelstein effect and spin-orbit torque in collinear ferromagnets.

Under linear response, the electric-field induced spin density $\delta\boldsymbol{S}$, known as the Edelstein effect[2], and the spin torque $\boldsymbol{T}$ can be expressed as: $\delta S^i = \chi^i_j E_j, T^i = t^i_j E_j$ ,where index $i$ denotes spin-polarization, $j$ denotes electric field directions. $\chi^i_j$ and $t^i_j$ are the Edelstein tensor and torkance, respectively. The Edelstein tensor can be expanded on the increasing orders of magnetic order parameter $\boldsymbol{m}$ in collinear ferromagnets by using non-magnetic point group symmetry as follows[3]:

$$\chi^i_j(\boldsymbol{m}) = \chi^{i(0)}_j + \chi^{i(1)}_{j,k} m_k + \chi^{i(2)}_{j,kl} m_k m_l + ...,$$

$$\chi^{i(n)}_{j,kl...} = \det(R)^{n\text{-}1} R_{ii'} R_{jj'} R_{kk'} R_{ll'} \chi^{i\ (n)}_{j',k'l'...}$$

Where, $R$ is the non-magnetic point group, det($R$) is the matrix determinant. Einstein summation rule is applied for repeated index. By doing this, the $T$-even and $T$-odd terms naturally map onto coefficients $\chi^{i(n)}_{j,kl...}$ with $n$ is even or odd numbers. In magnets, the spin torque can be seen as the consequence of the exchange interaction between spin-density $\delta\boldsymbol{S}$ and local magnetization. Therefore, the induced spin density $\delta\boldsymbol{S}$ will produce effective magnetic field and spin-torque on magnetization $\boldsymbol{M}$ via exchange interaction as: $\boldsymbol{B_E}=-J\delta\boldsymbol{S}/M_s$, $\boldsymbol{T}=\boldsymbol{M}\times\boldsymbol{B_E}$.

By introducing three SOC vectors $\boldsymbol{O}^a$ ($a$=1, 2, 3)[7], the Edelstein tensors can be expanded on the increasing orders of SOC strength as follows:

$$\chi_j^i = \chi_j^{i(0)} + \alpha_{j,k}^{i,a} O_k^a + \beta_{j,kl}^{i,ab} O_k^a O_l^b + ..$$

where $\chi_j^{i(0)}$ is the non-relativistic Edelstein tensor, $\boldsymbol{\alpha}$ and $\boldsymbol{\beta}$ are the rank-4 and rank-6 tensor coefficients, proportional to the first order and second order of SOC strength $\xi$, *i.e.* $\alpha\sim\xi$, $\beta\sim\xi^2$, respectively. The superscripts ($i$, $a$, $b$) are the spin-space indices, ($k$, $l$) are the orbital indices, $j$ is the applied electric field direction.

This approach also naturally yield $\boldsymbol{m}$ dependence of Edelstein tensors and torkances and can be further separated into time-reversal-even ($T$-even) and time-reversal-odd ($T$-odd) components, which satisfy $\chi_j^i(\boldsymbol{m}) = \chi_j^i(-\boldsymbol{m}), t_j^i(\boldsymbol{m}) = t_j^i(-\boldsymbol{m})$ and $\chi_j^i(\boldsymbol{m}) = -\chi_j^i(-\boldsymbol{m}), t_j^i(\boldsymbol{m}) = -t_j^i(-\boldsymbol{m})$, respectively. Under spin-point group operation $\{U||R\}$, where $U$ and $R$ are the symmetry operations in spin and real space, the non-relativistic Edelstein tensor $\chi_j^{i(0)}$, and the $\alpha$ and $\beta$ tensor coefficients can be determined from the transformations according to spin-group symmetry operations as follows:

(a) The nonrelativistic Edelstein tensors (zero order at SOC):

$$\chi_j^{i(0),\mathrm{odd}} = U_{ii'} R_{jj'} \chi_{j'}^{i'(0),\mathrm{odd}}; \quad \chi_j^{i(0),\mathrm{even}} = \det(U) U_{ii'} R_{jj'} \chi_{j'}^{i'(0),\mathrm{even}}$$

(b) The first-order SOC coefficients $\alpha$:

$$T\text{-odd:} \quad \alpha_{j,k}^{i,a} = U_{ii'} U_{aa'} R_{jj'} \det(U) \det(R) R_{kk'} \alpha_{j',k'}^{i',a'};$$

$$T\text{-even:} \quad \alpha_{j,k}^{i,a} = U_{ii'} U_{aa'} R_{jj'} \det(R) R_{kk'} \alpha_{j',k'}^{i',a'}$$

(c) The second-order SOC coefficients $\beta$:

$$T\text{-odd:} \quad \beta_{j,kl}^{i,ab} = U_{ii'} U_{aa'} U_{bb'} R_{jj'} R_{kk'} R_{ll'} \beta_{j',k'l'}^{i',a'b'}$$

$$T\text{-even:} \quad \beta_{j,kl}^{i,ab} = \det(U) U_{ii'} U_{aa'} U_{bb'} R_{jj'} R_{kk'} R_{ll'} \beta_{j',k'l'}^{i',a'b'}$$

In general, the spin group can be decomposed into the direct product $G_S = G_{NS} \times G_{SO}$, where $G_{NS}$ denotes the non-trivial spin group and $G_{SO}$ the spin-only group. For collinear ferromagnets, the spin-only group comprised the elements $\{C_{2n}T||E\}$ and $\{C_\varphi||E\}$. Here, $C_{2n}$ is a twofold spin rotation about an axis perpendicular to the spin quantization axis, $C_\varphi$ is an arbitrary spin rotation by angle $\varphi$ around the same axis, $T$ denotes the time reversal operation, and $E$ is the identity operation in real space. The non-trivial spin group for collinear ferromagnets is identical to the corresponding non-magnetic point group.

In the nonrelativistic limit of vanishing SOC, the spin-only constraints of collinear magnets require all nonrelativistic $T$-even Edelstein tensors to vanish identically. In contrast, the induced $T$-odd spin density may retain finite, with its spin-polarization aligned parallel to the magnetization. Consequently, the nonrelativistic Edelstein effect in collinear ferromagnets cannot exert a spin torque on the magnetization. It is solely through SOC-induced spin-group symmetry breaking that spin components transverse to the magnetization emerge, giving rise to SOT in collinear ferromagnets.

For the $C_{4v}$ point group, we first impose the spin-only constraints on the first-order SOC Edelstein tensor coefficients, yielding the nonzero $T$-odd $\alpha$ coefficients $\alpha_{j,k}^{1,2} = -\alpha_{j,k}^{2,1}$, and $T$-even coefficients $\alpha_{j,k}^{1,1} = \alpha_{j,k}^{2,2}, \alpha_{j,k}^{3,3}$. For the Ti/Ni(001) and Pt/CoFe(001) bilayers considered in this work, the non-trivial spin point group is $C_{4v}$, generated by two symmetry operations: $C_{4z}$, a four-fold rotation about the $z$ axis, and $M_y$, a mirror perpendicular to the $y$ axis. Upon further imposing the nontrivial spin-group symmetry constraints, the nonzero $T$-odd coefficients reduce to a single independent parameter $a_s$, given by: $\alpha_{y,1}^{1,2} = -\alpha_{x,2}^{1,2} = -\alpha_{y,1}^{2,1} = \alpha_{x,2}^{2,1} = a_s$ . The corresponding Edelstein tensors, with spin-polarization transverse to the magnetization $\boldsymbol{m}$, then take the form:

$$\chi_x^{1(1),\mathrm{odd}} = \alpha_{x,2}^{1,2} O_2^2 = -a_s O_2^2; \;\; \chi_y^{2(1),\mathrm{odd}} = \alpha_{y,1}^{2,1} O_1^1 = -a_s O_1^1$$

$$\chi_y^{1(1),\mathrm{odd}} = \alpha_{y,1}^{1,2} O_1^2 = a_s O_1^2; \;\; \chi_x^{2(1),\mathrm{odd}} = \alpha_{x,2}^{2,1} O_2^1 = a_s O_2^1$$

In the crystal frame, the vector form of the Edelstein tensors as a function of magnetization direction $\boldsymbol{m}$, reads:

$$\boldsymbol{\chi}_x^{(1),\mathrm{odd}} = \chi_x^{1(1),\mathrm{odd}} \boldsymbol{O}^1 + \chi_x^{2(1),\mathrm{odd}} \boldsymbol{O}^2 = a_s (O_2^1 O_1^2 - O_2^2 O_1^1, 0, O_2^1 O_3^2 - O_2^2 O_3^1) = a_s(-m_z, 0, m_x)$$

$$\boldsymbol{\chi}_y^{(1),\mathrm{odd}} = \chi_y^{1(1),\mathrm{odd}} \boldsymbol{O}^1 + \chi_y^{2(1),\mathrm{odd}} \boldsymbol{O}^2 = a_s (0, O_2^1 O_1^2 - O_2^2 O_1^1, O_1^2 O_3^1 - O_1^1 O_3^2) = a_s(0, -m_z, m_y)$$

Therefore, the nonequilibrium spin density induced by an in-plane electric field $\boldsymbol{E}_{xy}$=($E_x$, $E_y$,0) takes the form:

$$\delta \boldsymbol{S}^{(1),\text{odd}} = \boldsymbol{\chi}_x^{(1),\text{odd}} E_x + \boldsymbol{\chi}_y^{(1),\text{odd}} E_y = a_s \boldsymbol{m} \times (\boldsymbol{E}_{xy} \times \boldsymbol{z})$$

From a symmetry perspective, the spin indices of the spin Hall conductivity and the Edelstein tensors are subject to identical spin-group constrains. Notably, the induced spin-polarization $\delta\boldsymbol{S}^{(1),\text{odd}}$ shares the same functional form as that of the magnetic spin Hall effect, corroborating that first-order SOC-induced SOTs originate from the spin Hall effect. The corresponding $T$-even torque, described by a single parameter $a_t$ and expressed in vector form as a function of the magnetization direction $\boldsymbol{m}$, reads:

$$\boldsymbol{T}^{(1),\text{even}} = a_t \boldsymbol{m} \times [\boldsymbol{m} \times (\boldsymbol{E}_{xy} \times \boldsymbol{z})]$$

We find that this $T$-even torque is identical to the conventional damping-like (DL) torque widely adopted in the literature[1]. Our spin-group analysis rigorously demonstrates that it emerges at first-order in SOC.

Similarly, the nonzero $T$-even Edelstein tensor coefficients reduce to two independent parameters $b_\text{s}$ and $c_\text{s}$, given by:

$$\alpha_{y,1}^{1,1} = -\alpha_{x,2}^{1,1} = \alpha_{y,1}^{2,2} = -\alpha_{x,2}^{2,2} = b_s, \alpha_{y,1}^{3,3} = -\alpha_{x,2}^{3,3} = c_s$$

The corresponding Edelstein tensors for spin-polarization in spin space are given by:

$$\chi_x^{1(1),\text{even}} = -b_s O_2^1; \chi_x^{2(1),\text{even}} = -b_s O_2^2; \chi_x^{3(1),\text{even}} = -c_s O_2^3$$

$$\chi_y^{1(1),\text{even}} = b_s O_1^1; \chi_y^{2(1),\text{even}} = b_s O_1^2; \chi_y^{3(1),\text{even}} = c_s O_1^3$$

$$\boldsymbol{\chi}_x^{(1),\text{even}} = -b_s O_2^1 \boldsymbol{O}^1 - b_s O_2^2 \boldsymbol{O}^2 - c_s O_2^3 \boldsymbol{O}^3 = (0, -b_s, 0) + (b_s - c_s) m_y \boldsymbol{m}$$

$$\boldsymbol{\chi}_y^{(1),\text{even}} = b_s O_1^1 \boldsymbol{O}^1 + b_s O_1^2 \boldsymbol{O}^2 + c_s O_1^3 \boldsymbol{O}^3 = (b_s, 0, 0) + (c_s - b_s) m_x \boldsymbol{m}$$

The nonequilibrium spin-density induced by an in-plane electric field can be expressed in vector form as:

$$\delta \boldsymbol{S}^{(1),\text{even}} = \boldsymbol{\chi}_x^{(1),\text{even}} E_x + \boldsymbol{\chi}_y^{(1),\text{even}} E_y = b_s (\boldsymbol{E}_{xy} \times \boldsymbol{z}) + c_s [(\boldsymbol{E}_{xy} \times \boldsymbol{m}) \cdot \boldsymbol{z}] \boldsymbol{m}$$

The first term of the induced spin-polarization $\delta\boldsymbol{S}^{(1),\text{even}}$ shares the same functional form as the conventional spin Hall effect, while the second term originates from the anomalous Hall effect. This corroborates that first-order SOC-induced SOTs in $C_{4v}$ point-group symmetric systems are dominated by the spin Hall mechanism. However, the second term in the $T$-even spin-density exerts no torque on the magnetization, as its spin-polarization is aligned parallel to the magnetization direction. Accordingly, the corresponding $T$-odd torque in vector form reads:

$$\boldsymbol{T}^{(1),\text{odd}} = b_t \boldsymbol{m} \times (\boldsymbol{E}_{xy} \times \boldsymbol{z}).$$

This $T$-odd torque corresponds to the conventional field-like (FL) torque. Our spin-group analysis rigorously demonstrates that it arises at first-order in SOC.

Following the same symmetry procedure, the electric field induced nonequilibrium spin-density and SOTs at second-order in SOC, parametrized by the magnetization direction $\boldsymbol{m}$, reads (see Supplementary Note 1 for the details):

$$\delta\boldsymbol{S}^{(2),\text{odd}} = a_s' \boldsymbol{m} \times (\boldsymbol{E}_{xy} \times \boldsymbol{z}) + a_{2s}' (\boldsymbol{m} \cdot \boldsymbol{E}_{xy}) \boldsymbol{z} + a_{1s}' m_z (\boldsymbol{m} \cdot \boldsymbol{E}_{xy}) \boldsymbol{m}$$

$$\boldsymbol{T}^{(2),\text{even}} = a_t' \boldsymbol{m} \times [\boldsymbol{m} \times (\boldsymbol{E}_{xy} \times \boldsymbol{z})] + a_{1t}' (\boldsymbol{m} \cdot \boldsymbol{E}_{xy})(\boldsymbol{m} \times \boldsymbol{z})$$

$$\delta\boldsymbol{S}^{(2),\text{even}} = b_s' (\boldsymbol{E}_{xy} \times \boldsymbol{z}) + c_s' [(\boldsymbol{E}_{xy} \times \boldsymbol{m}) \cdot \boldsymbol{z}] \boldsymbol{m} + (d_s' - c_s') m_z^2 (\boldsymbol{E}_{xy} \times \boldsymbol{z}) + d_s' m_z [(\boldsymbol{E}_{xy} \times \boldsymbol{m}) \cdot \boldsymbol{z}] \boldsymbol{z}$$

$$\boldsymbol{T}^{(2),\text{odd}} = b_t' \boldsymbol{m} \times (\boldsymbol{E}_{xy} \times \boldsymbol{z}) + c_t' m_z^2 [\boldsymbol{m} \times (\boldsymbol{E}_{xy} \times \boldsymbol{z})] + d_t' m_z [(\boldsymbol{E}_{xy} \times \boldsymbol{m}) \cdot \boldsymbol{z}] (\boldsymbol{m} \times \boldsymbol{z})$$

As evident from the above spin-group expressions for the $C_{4v}$ point group, both the conventional DL and FL torques receive second-order SOC contribution. Moreover, additional torques beyond the conventional DL and FL forms appear at second-order in SOC.

## III. The first-principles calculations

To calculate the Edelstein tensors and spin-orbit torkances in practical materials, we employed first-principles computations based on the Korringa-Kohn-Rostoker (KKR) Green's function formalism[8][9]. The exchange and correlation potential is treated by local spin density approximation (LSDA) as parameterized by Vosko et al[10][11], while the potentials are represented using atomic sphere approximations. In the self-consistent calculations, angular momentum expansion cutoff of $l_{\max}$=3 and a $\boldsymbol{k}$-point mesh of 28×28×4 have been employed. The Edelstein tensors and torkances of Ti(4 MLs)/Ni(4 MLs) and Pt(6 MLs)/CoFe(3 MLs) bilayers are calculated through the linear response Kubo-Bastin formula. The torkances have been evaluated on a ***k***-point mesh of 200×200×1 in the Brillouin zone by considering lattice vibration at 300 K via alloy analogy model[12].

*Soc induced in-plane spin-texture and the Edelstein effect from the band structure.* We first elucidate the band structure origin of the SOC-induced Edelstein effect in collinear ferromagnets. As depicted in Fig. 1 (a), the Ti/Ni and Pt/CoFe bilayers are constructed on the bcc(001) lattice with a lattice constant $a$=5.416 $a_0$ ($a_0$ is the Bohr radius). The nonmagnetic crystal structure belongs to point group 4*mm* ($C_{4v}$). The magnetic moment is aligned along $z$ direction, giving rise to a corresponding magnetic point group (MPG) of $4m'm'$. This MPG is generated by two symmetry operations: $C_{4z}$ (four-fold rotation about $z$ axis) and $M_yT$ (the combination of mirror plane perpendicular to $y$ axis and time reversal). In the absence of SOC, the spin-texture of the bands possesses only a $z$-component collinear with the magnetization. Upon inclusion of SOC, however, in-plane spin components ($S_x$, $S_y$) transverse to the magnetization emerge. The symmetry constraints imposed by the MPG on the ***k***-resolved spin-textures at the Fermi energy are illustrated in Fig.1(d) and (e), and they adhere to the following restrictions:

$$C_{4z}\text{: } (k_x, k_y; S_x, S_y, S_z)\rightarrow(k_y, -k_x; S_y, -S_x, S_z);$$

$$M_yT\text{: } (k_x, k_y; S_x, S_y, S_z)\rightarrow(-k_x, k_y; S_x, -S_y, S_z).$$

Notably, the Pt/CoFe bilayer exhibits a more elaborate and richer spin texture than the of Ti/Ni bilayer. Simultaneously, the in-plane spin components $S_x$ and $S_y$ exhibit a relatively uniform spatial distribution in Ti/Ni, whereas those in Pt/CoFe are markedly more localized around the Γ point.

Within the framework of linear response and perturbation theory, the $T$-even and $T$-odd Edelstein effect (corresponding to the $T$-odd and $T$-even torques) originate predominantly from intraband and interband processes, respectively[1]. For instance, under the constant-relaxation-time and low-temperature approximation, the electric field induced $T$-even spin-density reads:

$$\delta \boldsymbol{S}^{\text{even}} = \tau \sum_{n,\boldsymbol{k}} \text{Re}[<n,\boldsymbol{k} \,|\, e\boldsymbol{E}\cdot\hat{\boldsymbol{v}} \,|\, n,\boldsymbol{k}><n,\boldsymbol{k} \,|\, \hat{\boldsymbol{S}} \,|\, n,\boldsymbol{k}>] \frac{\partial}{\partial \varepsilon} f_{n,\boldsymbol{k}}$$

where $\tau$ is the constant relaxation time, $n$ the band index, ***k*** the crystal momentum, $e$ the elementary charge, $\boldsymbol{E}$ the applied electric field, and $f_{n,\boldsymbol{k}}$ the Fermi-Dirac distribution function. $\hat{\boldsymbol{v}}$ and $\hat{\boldsymbol{S}}$ denote the velocity and spin operators, respectively. The associated *T*-even Edelstein tensor $\chi_j^{i,\text{even}}$ is accordingly given by:

$$\chi_j^{i,\text{even}} \approx -Ee\tau \sum_{\boldsymbol{k}} v_j(\boldsymbol{k}) S_i(\boldsymbol{k}) \delta(\varepsilon_F - \varepsilon_{\boldsymbol{k}})$$

where $\boldsymbol{v}(\boldsymbol{k})$ is the Fermi velocity and $\boldsymbol{S}(\boldsymbol{k})$ is the $\boldsymbol{k}$-resolved spin-texture. Broken inversion symmetry is a prerequisite for a nonvanishing Edelsten effect. In inversion-symmetric systems, the relations $\boldsymbol{v}(-\boldsymbol{k})=-\boldsymbol{v}(\boldsymbol{k})$ and $\boldsymbol{S}(-\boldsymbol{k})=\boldsymbol{S}(\boldsymbol{k})$ render the integrand of the Edelstein tensor an odd function of $\boldsymbol{k}$, which integrates to zero over the full Brillouin zone. In total, 21 non-magnetic point groups with broken inversion symmetry can support a finite Edelstein effect in collinear ferromagnets[1]. It follows that the SOC induced *T*-even transverse spin texture ($S_x$, $S_y$) directly accounts for the *T*-odd Spin-orbit torque, i.e., the Rashba torque. This contribution is governed primarily by Fermi-surface properties, and analogous to electric conductivity, scales linearly with the relaxation time[13]. Correspondingly, the SOC induced *T*-odd Edelstein effect underpins the *T*-even SOT, i.e., the spin Hall torque[14][15].

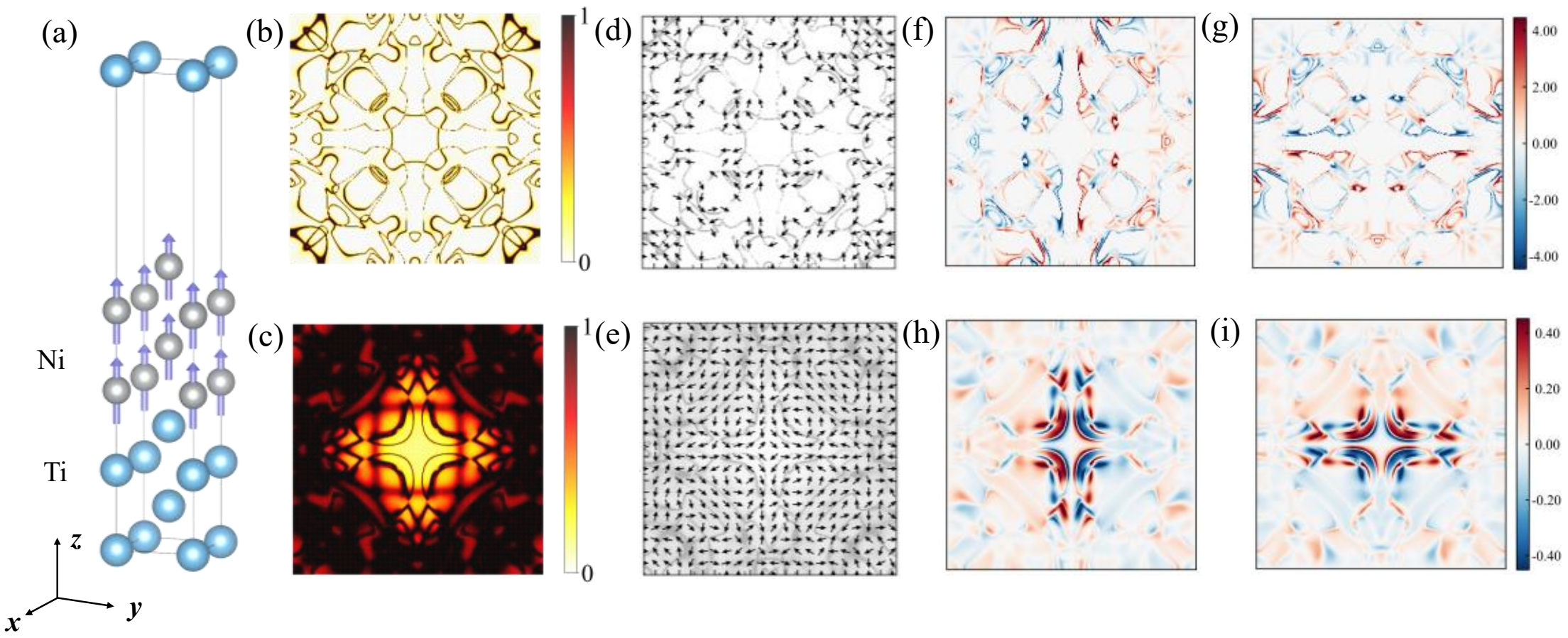

Fig. 1. (a) The crystal structures of $Ti_4/Ni_4(001)$ bilayers with magnetic moments along the *z* direction. (b) and (c) are the $\boldsymbol{k}$-resolved Bloch spectra of $Ti_4/Ni_4(001)$ and $Pt_6/CoFe_3(001)$ bilayers at Fermi energy, respectively. (d) and (e) are the corresponding SOC induced $\boldsymbol{k}$-resolved in-plane spin texture ($S_x$, $S_y$). (f) and (g) show the color-coded $\boldsymbol{k}$-resolved in-plane spin components $S_x$ and $S_y$ at the Fermi level for $Ti_4/Ni_4(001)$ bilayer. (h) and (i) present the corresponding distributions for $Pt_6/CoFe_3(001)$ bilayers, respectively.

*Edelstein tensor and torkance as a function of SOC strength.* We then first investigate the SOC scaling of the Edelstein effect and spin-orbit torque in the orbit-torque-dominated Ti/Ni bilayer and spin-Hall-torque-dominated Pt/CoFe bilayer[6], with the magnetization aligned along the *z* direction. The SOC strength is scaled in our calculations by varying the speed of light. The nonzero Edelstein tensors and torkances for ***m***//*z* and 4*m*'*m*' MPG are: *T*-odd $\chi_x^x = \chi_y^y$, $t_x^x = t_y^y$, and *T*-even: $\chi_x^y = -\chi_y^x, t_x^y = -t_y^x$.

As shown in Fig. 2, for both bilayer systems, the induced spin density and SOT vanish in the absence of SOC, confirming that SOC is indispensable for spin-torque in collinear ferromagnetic bilayers, regardless the underlying spin or orbital Hall mechanism. For the Ti/Ni system, both the induced Edelstein effect and the resulting torques grow monotonically with increasing SOC strength. In contrast, for Pt/CoFe bilayer the *T*-odd Edelstein tensors $\chi_x^x$, $\chi_y^x$, and the torkance $t_x^x$ first rise and then decline with increasing SOC strength, whereas the DL torkance $t_y^x$ increase monotonically. The monotonic SOC dependence of the DL torkance in Pt/CoFe originates from the enhanced weight of the second-order SOC contribution in the strong SOC system, as we elaborate in subsequent sections.

Formally, the Edelstein tensors and torkances can be written as increasing power of SOC scaling as follows:

$$\chi_x^x = \chi_x^{x(1)} + \chi_x^{x(2)} + \ldots = \lambda_{xx}^{(1)}(\xi/\xi_0) + \lambda_{xx}^{(2)}(\xi/\xi_0)^2 + \ldots$$

$$\chi_y^x = \chi_y^{x(1)} + \chi_y^{x(2)} + \ldots = \lambda_{xy}^{(1)}(\xi/\xi_0) + \lambda_{xy}^{(2)}(\xi/\xi_0)^2 + \ldots$$

$$t_x^x = t_x^{x(1)} + t_x^{x(2)} + \ldots = \eta_{xx}^{(1)}(\xi/\xi_0) + \eta_{xx}^{(2)}(\xi/\xi_0)^2 + \ldots$$

$$t_y^x = t_y^{x(1)} + t_y^{x(2)} + \ldots = \eta_{xy}^{(1)}(\xi/\xi_0) + \eta_{xy}^{(2)}(\xi/\xi_0)^2 + \ldots$$

As depicted in Fig. 2, for both bilayer heterostructures, all Edelstein tensors and torkances are well reproduced by an SOC expansion truncated at second order. The corresponding fitting parameters $\lambda$ and $\eta$ are tabulated in Table I. In the orbital-Hall-dominated Ti/Ni bilayer, the first-order Edelstein tensors are substantially larger in magnitude than their second-order coefficients. This hierarchy confirms that the SOC-driven response is overwhelmingly dominating by the first-order contribution in the weak SOC case. By contrast, in the spin Hall dominated Pt/CoFe system, the second-order coefficients of the Edelstein tensors are comparable in magnitude to their first-order counterparts, indicating a substantial second-order SOC contribution in the strong SOC system.

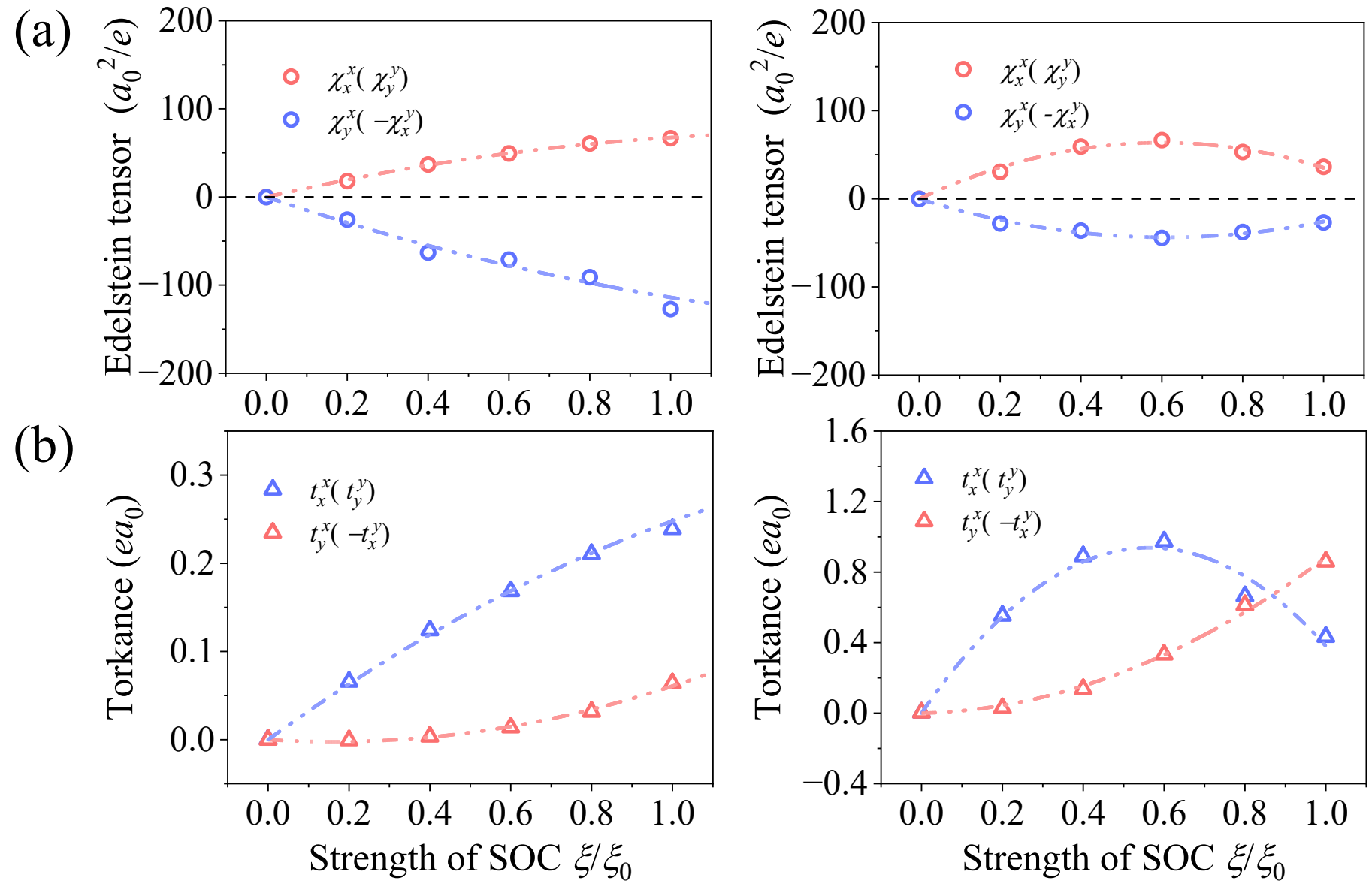

Fig. 2. (a) The calculated Edelstein tensors $\chi_x^x(\chi_y^y)$, $\chi_y^x(-\chi_x^y)$ and (b) torkances $t_x^x(t_y^y)$, $t_y^x(-t_x^y)$ as a function of SOC scaling $\xi/\xi_0$ for Ti/Ni (left panels) and Pt/CoFe (right panels) bilayers, respectively, with magnetization aligned along the $z$ direction. $\xi_0$ denotes the intrinsic SOC strength. Open symbols represent the first-principles results, while dashed lines show the corresponding fitting curves. Matched Edelstein tensor-torkance pairs $\chi_x^x(\chi_y^y)$ and $t_x^y(-t_y^x)$, $\chi_x^y(-\chi_y^x)$ and $t_x^x(t_y^y)$ are plotted in consistent colors: blue for the $T$-even Edelstein tensor and $T$-odd torque, and the red for the $T$-odd Edelstein tensor and $T$-even torque.

Table I. SOC scaling parameters for Ti/Ni(001) and Pt/CoFe(001) bilayer with $\boldsymbol{m}//z$. The SOC expansion coefficients of Edelstein tensor and torkance are given in units of $a_0^2/e$ and $ea_0$, respectively.

| | Edelstein Tensors | | Spin-orbit torkances | |
|---|---|---|---|---|
| | $T$-odd | $T$-even | FL | DL |

| coefficients | $\lambda_{xx}^{(1)}, \lambda_{xx}^{(2)}$ | $\lambda_{xy}^{(1)}, \lambda_{xy}^{(2)}$ | $\eta_{xx}^{(1)}, \eta_{xx}^{(2)}$ | $\eta_{xy}^{(1)}, \eta_{xy}^{(2)}$ |
|---|---|---|---|---|
| Ti/Ni | 104.86, -37.40 | -153.46, 39.51 | 0.331, -0.083 | -0.0286, 0.0892 |
| Pt/CoFe | 212.08, -176.89 | -143.06, 117.26 | 3.33, -2.95 | 0.055, 0.828 |

Turning to the torkance coefficients, as listed in Table I the first-order SOC coefficient of the *T*-odd torque in Ti/Ni is markedly larger than its second-order coefficient, indicating that the *T*-odd torque in Ti/Ni scales linearly with SOC strength. For the *T*-even torque, by contrast, the second-order SOC coefficient exceeds the first-order value, indicating that higher-order SOC contributions play a more prominent role in the *T*-even torque. For Pt/CoFe, both the FL and DL torques exhibit substantial second-order SOC contributions since it has much stronger SOC strength.

*The relation between layer-resolved Edelstein effect and torque.* To gain deeper insight into the Edelstein effect and the associated spin-orbit torques, we present layer-resolved results for the Ti/Ni bilayer with ***m***//*z* in Fig. 3 (The corresponding layer-resolved plots for Pt/CoFe are provided in the Supplementary Material Note 2). The relation between the layer-resolved spin density and torque can be described qualitatively as $\boldsymbol{T}_i = \frac{-J_i V_i}{\mu_i}(\boldsymbol{m} \times \delta \boldsymbol{S}_i)$, where $i$ denotes the layer index, and $\mu_i$, $V_i$, $J_i$ correspond to the magnetic moment, effective volume and exchange energy of layer $i$, respectively. $\delta\boldsymbol{S}_i$ and $\boldsymbol{T}_i$ represent the corresponding induced spin density and torque. Consequently, $\chi_x^x(\chi_y^y)$ and $t_x^y(-t_y^x)$, $\chi_x^y(-\chi_y^x)$ and $t_x^x(t_y^y)$ exhibit a one-to-one correspondence.

Specifically, for Ti/Ni bilayers, the spin density in all layers increase monotonically with increasing SOC strength, consistent with the monotonic growth of Edelstein tensor shown in Fig. 2 (a). It is also noteworthy that the SOC induced spin density in the Ti layer is sizable, despite Ti being a light metal with negligible spin Hall conductivity. We attribute this behavior to orbital-to-spin conversion in Ni. Specifically, the orbital current generated in Ti via the OHE is converted into a spin current at the Ti/Ni interface: one portion transmits across the interface to exert a torque on Ni, while the remaining part backflows into the Ti layer and forms the observed spin density. This mechanism gives rise to a spin density tensor in Ti with a magnitude comparable to that in the Pt layer of the Pt/CoFe heterostructure (see Supplementary Note 2). As shown in Fig. 2 (b), since magnetic moments in the nonmagnetic Ti and Pt layers are nearly zero, the layer-resolved torkance is almost entirely confined to the magnetic layers. For Ti/Ni, the net torques arise from the asymmetric torque distribution across the Ni layers, whereas for Pt/CoFe, the net torques predominantly originate from the CoFe layer adjacent to the heavy metal Pt.

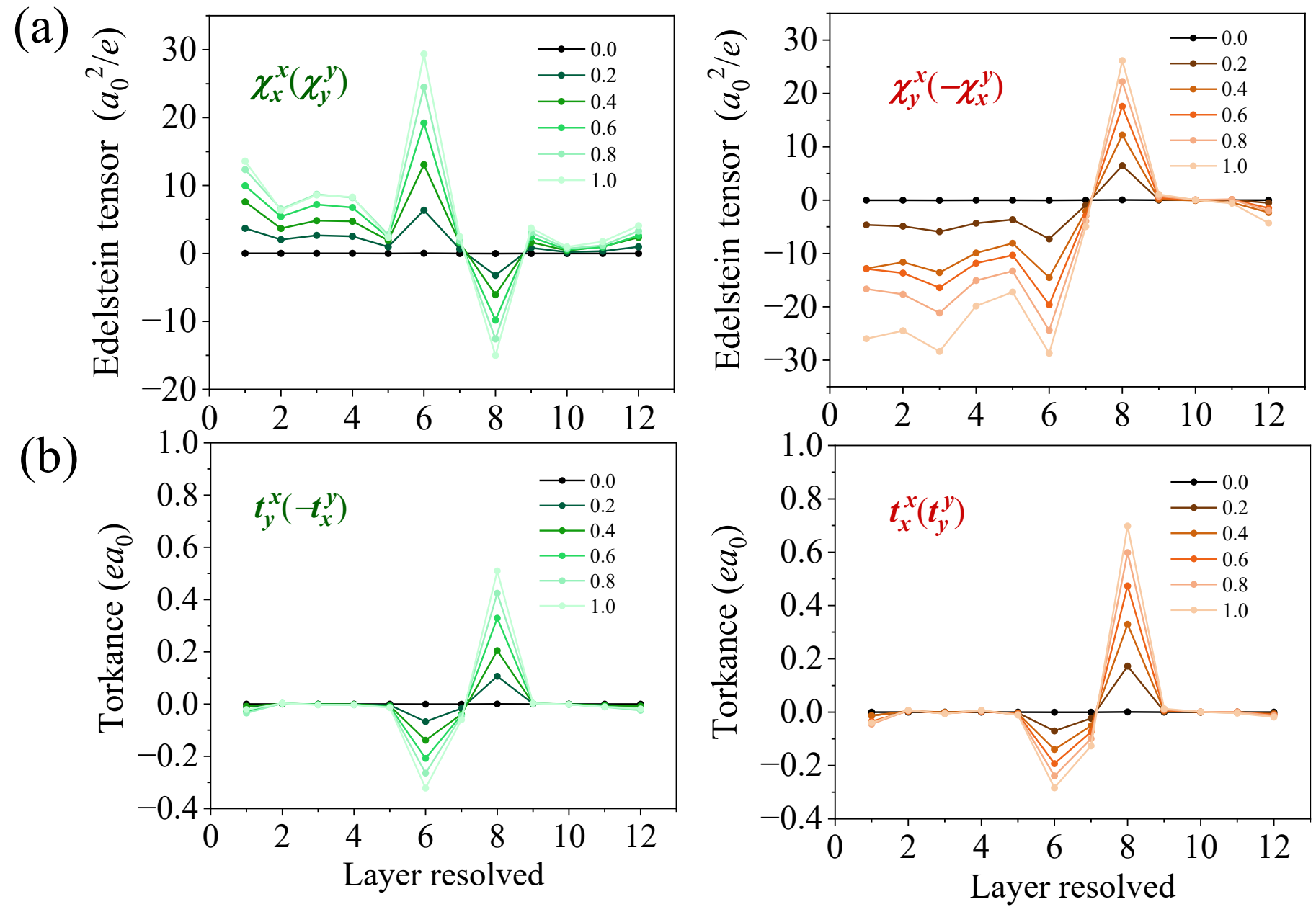

Fig. 3. (a) Layer-resolved Edelstein tensors $\chi_x^x(\chi_y^y)$, $\chi_x^y(-\chi_y^x)$, and (b) corresponding torkances $t_x^y(-t_y^x)$, $t_x^x(t_y^y)$ for the Ti/Ni bilayer, plotted as a function of the SOC scaling factor $\xi/\xi_0$ ranging from 0 to 1.

*Magnetization direction **m** dependence of Edelstein effect and SOT.* To validate the symmetry-derived tensor forms for spin density and torkance, we perform angular-dependent calculations on Edelstein tensors and torkances in Ti/Ni(001) and Pt/CoFe(001). The analytical expressions deduced from the spin-group theory up to second-order SOC for $C_{4v}$ point group are summarized as follows:

$$\delta \boldsymbol{S}^{\text{odd}} = (a_s + a_s^{'})\boldsymbol{m} \times (\boldsymbol{E}_{xy} \times \boldsymbol{z}) + a_{2s}^{'}(\boldsymbol{m} \cdot \boldsymbol{E}_{xy})\boldsymbol{z} + a_{1s}^{'} m_z (\boldsymbol{m} \cdot \boldsymbol{E}_{xy})\boldsymbol{m}$$

$$\delta \boldsymbol{S}^{\text{even}} = (b_s + b_s')(\boldsymbol{E}_{xy} \times \boldsymbol{z}) + (c_s + c_s')[(\boldsymbol{E}_{xy} \times \boldsymbol{m}) \cdot \boldsymbol{z}]\boldsymbol{m} + (d_s' - c_s')m_z^2(\boldsymbol{E}_{xy} \times \boldsymbol{z}) + d_s' m_z[(\boldsymbol{E}_{xy} \times \boldsymbol{m}) \cdot \boldsymbol{z}]\boldsymbol{z}$$

$$\boldsymbol{T}^{\text{even}} = (a_t + a_t')\boldsymbol{m} \times [\boldsymbol{m} \times (\boldsymbol{E}_{xy} \times \boldsymbol{z})] + a_{1t}'(\boldsymbol{m} \cdot \boldsymbol{E}_{xy})(\boldsymbol{m} \times \boldsymbol{z})$$

$$\boldsymbol{T}^{\text{odd}} = (b_t + b_t')\boldsymbol{m} \times (\boldsymbol{E}_{xy} \times \boldsymbol{z}) + c_t' m_z^2[\boldsymbol{m} \times (\boldsymbol{E}_{xy} \times \boldsymbol{z})] + d_t' m_z[(\boldsymbol{E}_{xy} \times \boldsymbol{m}) \cdot \boldsymbol{z}](\boldsymbol{m} \times \boldsymbol{z})$$

For the calculations in Ti/Ni bilayer, the magnetization has been constrained to in-plane (*xy* plane). The typical Edelstein tensors and torkances for magnetization within the *xy* plane are as follows, with $\phi$ being defined as the angle between $\boldsymbol{m}$ and *x*-axis, i.e. $m_x = \cos\phi, m_y = \sin\phi, m_z = 0$:

$$\chi_x^x = (c_s + c_s')m_x m_y, \chi_y^x = -(c_s + c_s')m_x^2 + (b_s + b_s'), \chi_x^z = -(a_s + a_s')m_x + a_{2s}' m_x$$

$$t_x^x = -(a_t + a_t' - a_{1t}')m_x m_y, t_y^x = -(a_t + a_t' - a_{1t}')m_y^2, t_x^z = -(b_t + b_t')m_x$$

For the Edelstein tensors shown in Fig. 4 (a) and torque tensors in Fig. 4 (b), the component $\chi_x^x$ and $t_x^x$ matches the $\sin 2\phi$ form, $\chi_y^x$ and $t_y^x$ shows a quadratic $\cos^2\phi$ term, $\chi_x^z$ and $t_x^z$ follow $\cos\phi$ dependence, respectively. The spin density components show satisfactory agreement between the calculation results and analytical equations, confirming that the angular dependence of Edelstein effect can be relatively accurately captured by the expansion up to second order SOC. For the Ti/Ni bilayer with in-plane magnetization, the *T*-even torque parameter (in the unit of $ea_0$) are found to be $a_t' + a_t - a_{1t}' = -0.0141$ ,which is one order of magnitude smaller than the conventional FL torque parameters $b_t + b_t' = 0.145$ . However, for components exhibiting pronounced deviations (e.g. $t_x^x$ and $t_y^x$), a higher order SOC expansion may be required.

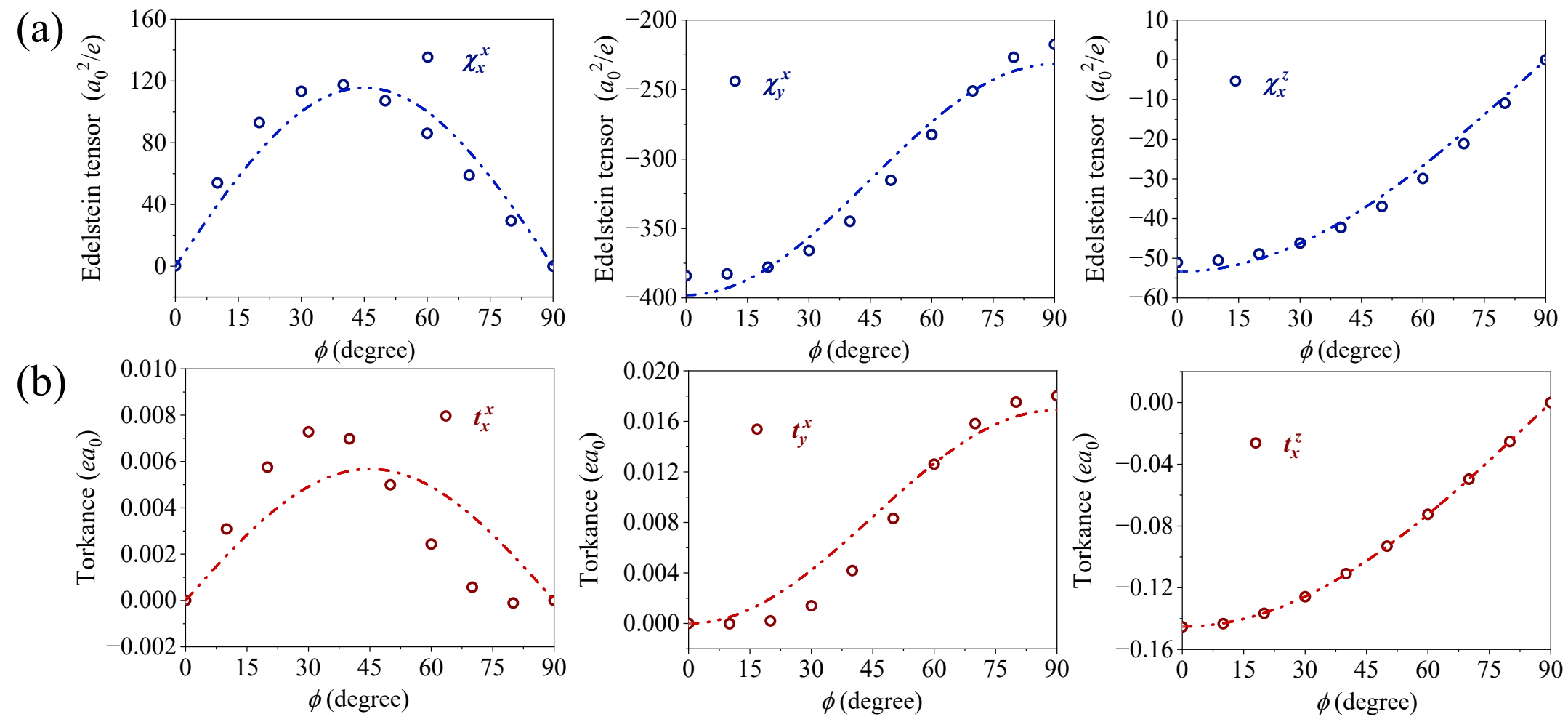

Fig. 4. (a) The angular dependence of Edelstein tensors $\chi_x^x, \chi_y^x, \chi_x^z$, and (b) torkances $t_x^x, t_y^x, t_x^z$ with magnetic moment rotating within *xy* plane in Ti/Ni(001) bilayer. $\phi$ is defined as the angle between the magnetization and *x*-axis. Open symbols represent the first-principles results, while dashed lines show the corresponding fitting curves according to analytic expressions.

For angular dependence calculations of the Pt/CoFe(001) bilayer, the magnetization is constrained to lie in the *xz* plane (out-of-plane geometry). The Edelstein tensors and torkances for magnetization in the *xz* plane are given below, where $\theta$ is defined as the angle between the magnetization and *z*-axis, i.e. $m_x = \sin\theta, m_y = 0, m_z = \cos\theta$:

Representative *T*-odd Edelstein tensors and *T*-even torkances for ***m*** confined to the *xz* plane:

$$\chi_x^x = (a_s + a_s^{'})m_z + a_{1s}^{'} m_x^2 m_z; \quad \chi_y^y = (a_s + a_s^{'})m_z; \chi_x^z = -(a_s + a_s^{'} - a_{2s}^{'})m_x + a_{1s}^{'} m_z^2 m_x;$$

$$t_x^y = (a_t + a_t^{'} - a_{1t}^{'})m_x^2 + (a_t + a_t^{'})m_z^2; t_y^x = -(a_t + a_t^{'})m_z^2; t_y^z = (a_t + a_t^{'})m_x m_z$$

Representative *T*-even Edelstein tensors and *T*-odd torkances for ***m*** confined to the *xz* plane:

$$\chi_x^y = -(b_s + b_s^{'}) - (d_s^{'} - c_s^{'})m_z^2; \quad \chi_y^x = (b_s + b_s^{'}) - (c_s + c_s^{'})m_x^2 + (d_s^{'} - c_s^{'})m_z^2$$

$$t_x^x = (b_t + b_t^{'})m_z + c_t m_z^3; \quad t_y^y = (b_t + b_t^{'})m_z + c_t m_z^3 + d_t^{'} m_z m_x^2; t_x^z = -(b_t + b_t^{'})m_x - c_t m_x m_z^2;$$

As shown in Fig. 5, the calculated angular dependence of both the Edelstein tensors and torkances are well reproduced by the analytic expressions derived above, corroborating the validity of the spin-group symmetry analysis. The conventional DL torkance parameters are found to be $a_t + a_t^{'} = -0.885$. By contrast, the substantially weaker DL torkance component originates from second-order SOC: $a_{1t}^{'} = -0.0619$. The corresponding conventional FL torkance parameters read: $b_t + b_t^{'} = 0.569$, while the additional FL contribution arising from second-order SOC with comparable magnitude is given by: $c_t^{'} = -0.137$, $d_t^{'} = -0.215$.

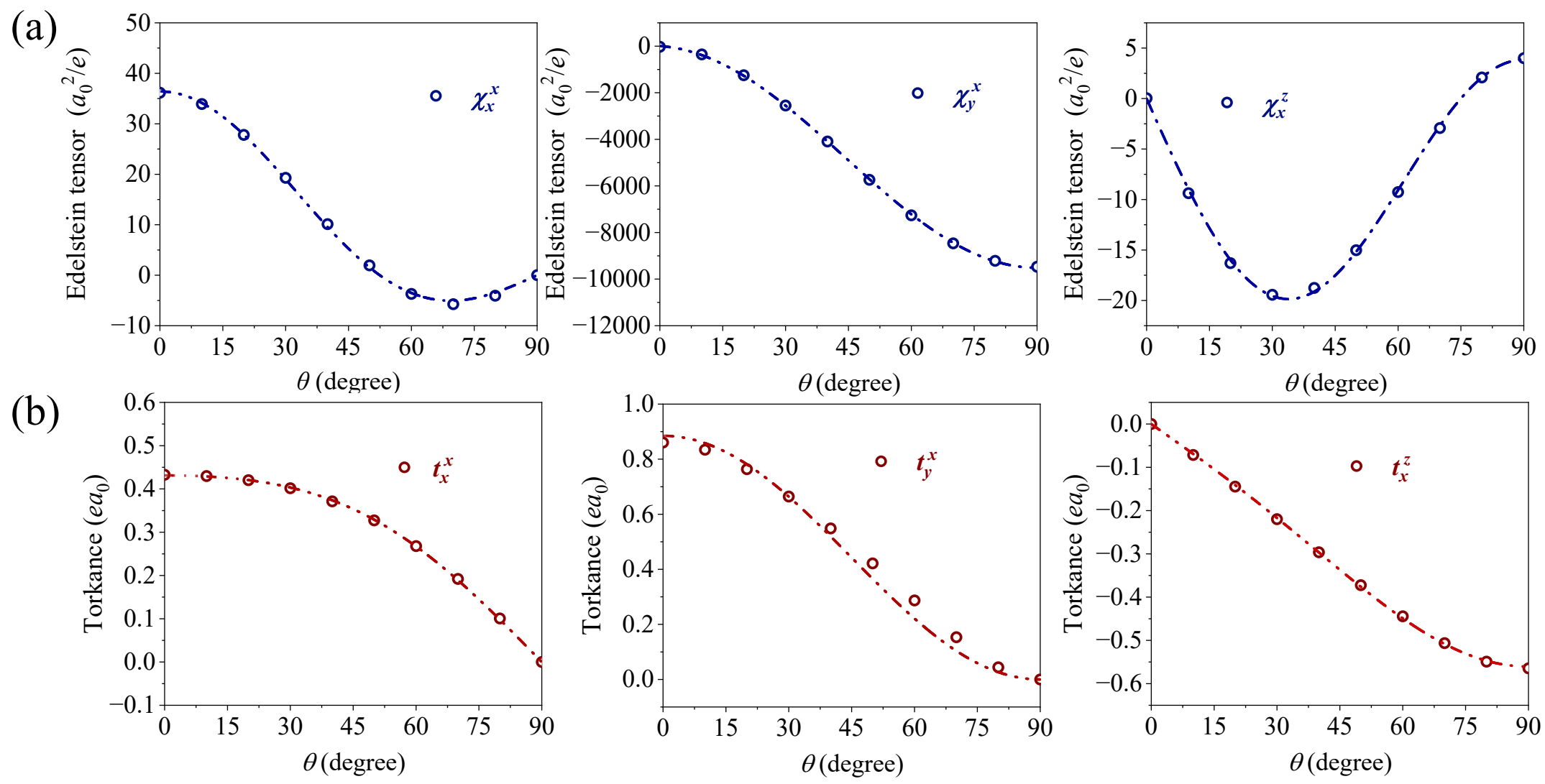

Fig. 5. (a) Angular dependence of the Edelstein tensors $\chi_x^x, \chi_y^x, \chi_x^z$, and (b) corresponding torkances $t_x^x, t_y^x, t_x^z$ in Pt/CoFe(001) bilayers, with the magnetization confined to the *xz* plane. $\theta$ is defined as the angle between the magnetization and *z*-axis. Open symbols represent the first-principles results, while dashed lines show the corresponding fitting curves according to analytic expressions.

For polycrystalline collinear ferromagnetic bilayers, the effective non-trivial spin-group reduces to the nonmagnetic point group $C_{\infty v}$. Up to second order in SOC, the functional forms of the induced spin-density and torques are identical to those derived for the $C_{4v}$ point group. For other noncentrosymmetric point groups, the spin-group constraints may give rise to distinct forms of the Edelstein effect and SOT. Upon further lowering of the crystal symmetry, additional contributions to the Edelstein effect and SOT emerge. For instance, the $C_{3v}$ point group describing fcc-Pt(111)/Co(111) bilayer yields the same first-order SOC Edelstein tensor and torkances (conventional form of DL and FL torques) as $C_{4v}$ point group, but differs in the second-order SOC contributions to both the Edelstein effect and SOT.

The $T$-even torques for $C_{3v}$ point group (with generators $C_{3z}$ and $M_y$) up to second-order SOC take the form:

$$\boldsymbol{T}^{\mathrm{even}} = (a_t + a_t^{'})\boldsymbol{m}\times[\boldsymbol{m}\times(\boldsymbol{E}_{xy}\times\boldsymbol{z})] + a_{1t}^{'}(\boldsymbol{m}\cdot\boldsymbol{E}_{xy})(\boldsymbol{m}\times\boldsymbol{z}) + a_{3m}^{'}\boldsymbol{m}\times[(m_x E_x - m_y E_y)\boldsymbol{x} - (m_y E_x + m_x E_y)\boldsymbol{y}]$$

The first two terms coincide with those derived for the $C_{4v}$ point group, whereas the third term constitute a characteristic torque unique to $C_{3v}$ point group, commonly referred to as the "3$m$ torque"[16][17]. To extract the key torque parameters for the Pt/Co(111) bilayer, we perform first-principles calculations with the magnetization confined to the $xz$ plane. The analytic expressions describing the angular dependence of the $T$-even torkances are given by:

$$t_y^x = -(a_t + a_t^{'})(m_y^2 + m_z^2) + a_{1t}^{'} m_y^2 + a_{3m}^{'} m_z m_x = -(a_t + a_t{'})\cos^2\theta + a_{3m}{'}\sin\theta\cos\theta$$

$$t_y^z = (a_t + a_t^{'})m_x m_z + a_{3m}^{'}(m_y^2 - m_x^2) = (a_t + a_t^{'})\sin\theta\cos\theta - a_{3m}^{'}\sin^2\theta$$

As shown in Fig. 6, the calculated results are in good agreement with the analytical expressions derived above, yielding the torkance parameters: $a_t + a_t{'} = -0.880$ , $a_{3m}^{'} = 0.104$. It is clear that the conventional DL torque is dominant and much larger than that of the 3$m$ torque arising from the second-order SOC. Field-free switching of a perpendicular magnetization (PMA), cannot be achieved by conventional torques alone. In what follows, we discuss the potential of second order SOC contributions to SOT for enabling such field-free switching of a PMA.

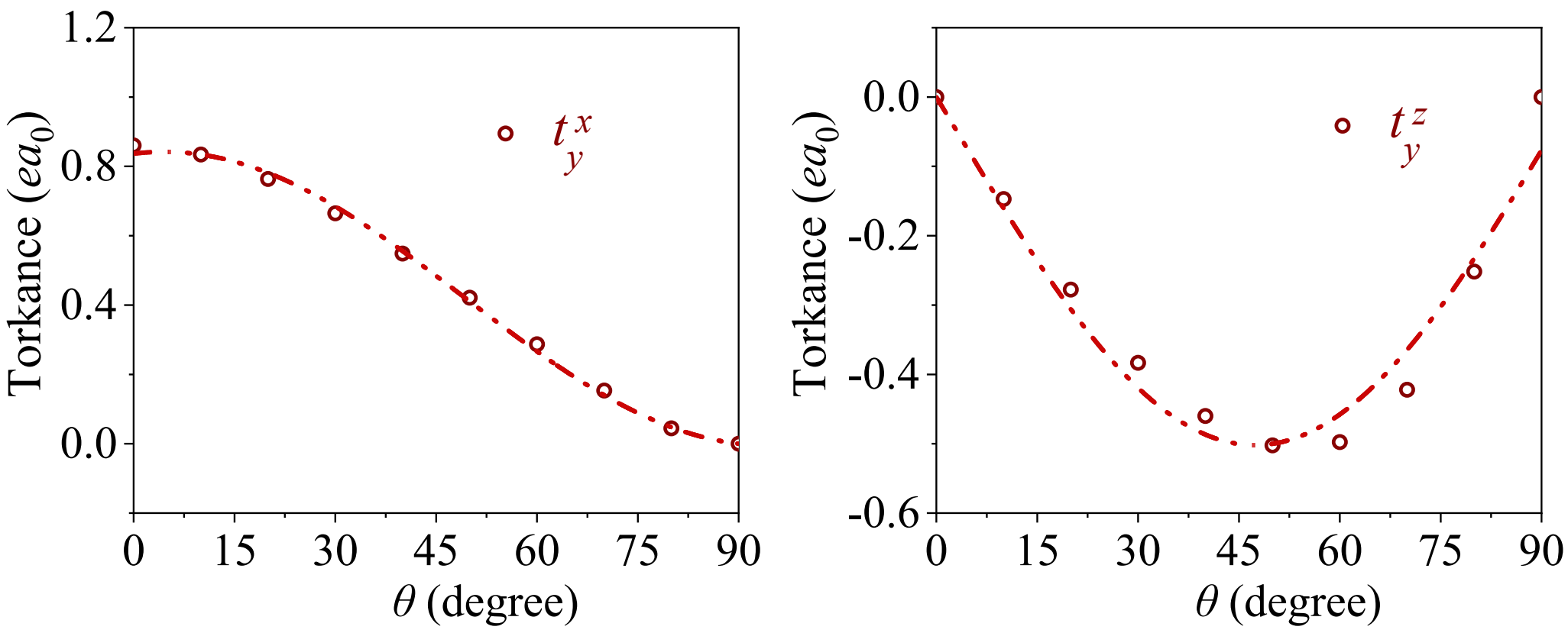

Fig. 6 Angular dependence of *T*-even torkances $t_y^x, t_y^z$ in Pt/Co(111) bilayers, together with corresponding fits to analytic expressions. The magnetization is confined to the *xz* plane, with $\theta$ is defined as the angle between the magnetization and the *z*-axis. Open symbols represent the first-principles results, while dashed lines show the corresponding fitting curves according to analytic expressions.

For thin film systems with $C_{\infty v}$ and $C_{4v}$ point group symmetry and uniaxial perpendicular magnetic anisotropy, an electric field applied along the *y* direction generates a sufficiently large conventional DL torque that rotates the initial magnetization ***m***//*z* towards the in-plane *x* direction. As a consequence, the second-order SOC *T*-even torque vanish in this in-plane configuration, precluding deterministic perpendicular magnetization switching. This behavior is qualitatively altered in $C_{3v}$ system, where the symmetry-allowed 3*m* torque provides an additional switching pathway. As illustrated in Fig. 7 (a), when the initial out-of-plane magnetization ***m***//*z* is rotated toward the *x* direction by conventional DL torque under an applied electric field $+E_y$, the resulting 3*m* *z*-torque develops a finite *z* component whose sign is determined by the sign of the 3*m* torques. In the Pt/Co(111) bilayer, an applied electric field $+E_y$ drives the magnetization from the initial +*z* state to final -*z* state, whereas $-E_y$ induces switching from the -*z* state to the +*z* state. This field-free switching scenario is corroborated by Landau-Lifshitz-Gilbert (LLG) simulation, as shown in Fig. 7 (b). A current density of $+J_y$=2.7×10$^8$ A/cm$^2$ drives deterministic 180° reversal of the initial magnetization from the +*z* to the -*z* orientation, and the magnetization switches back to +*z* state under a reversed current of $-J_y$=2.7×10$^8$ A/cm$^2$.

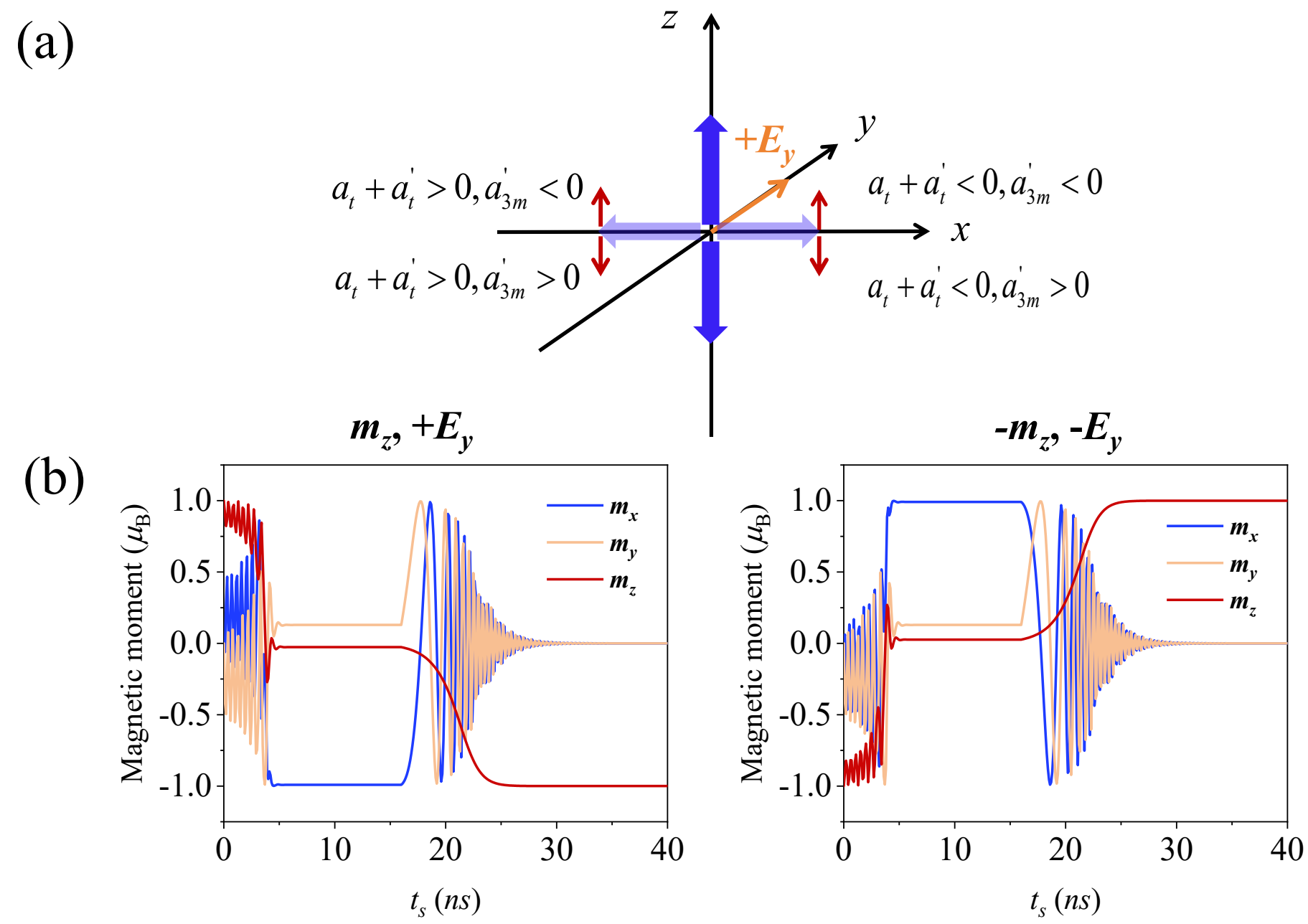

Fig. 7 (a). Schematic illustration of the $z$-torque driving field-free PMA switching under an applied electric filed $+E_y$. Blue arrows indicate the initial magnetization along the $+z$ or $-z$ directions, light blue arrows denote the magnetization along the $+x$ and $-x$ directions, and red arrows mark the direction of the $z$-torque for opposite signs of the conventional torque and the 3$m$ torque. (b). LLG simulations of field-free PMA switching in a system with 3$m$ point group symmetry. The perpendicular magnetic anisotropy field $B_{\mathrm{ani}}$ is set to 0.1 T. The simulation time is 40 *ns*, with a 16 *ns* current pulse applied.

Similarly, the viability of field-free perpendicular magnetization switching in thin film system with other noncentrosymmetric point group can be qualitatively evaluated from spin-group symmetry arguments. Apart from $C_{4v}$ and $C_{\infty v}$ point groups, the field-free switching of PMA is not favorable in other point groups with two perpendicular mirror planes or with a rotation axis $C_{nz}$ with $n \geq 4$ including $mm2$ ($C_{2v}$), 6$mm$, 4 and 6 point groups *etc*. In contrast, point groups with symmetry lower than $C_{3v}$, such as point group $m$ and 3, are compatible with field-free switching in principles. For certain other point groups, including 422, -42$m$, the $T$-odd and $T$-even torques cannot even be cast into the conventional DL or FL forms.

*Edelstein effect and SOT in PtMnSb.* We finally investigate PtMnSb, a collinear ferromgnets with broken inversion symmetry that crystallizes in the cubic -43$m$ point group. The non-trivial spin-group of bulk PtMnSb is generated by $S_{4z}$ and $C_{3[111]}$. Spin-symmetry constraints prohibit finite spin density and torques at zeroth and first order in SOC; consequently, no conventional FL and DL SOT can arise in this compound. Both $T$-odd and $T$-even Edelstein tensors and torkances instead emerge at second order in SOC. In line with this symmetry analysis, our first-principles calculations confirm that the Edelstein tensors and torkances in PtMnSb exhibit sizable third-order SOC contributions.

The spin-group symmetry imposed non-zero Edelstein tensors in matrix form can be written as follows:

$$\chi^{(2),\mathrm{odd}} = a_1'\begin{pmatrix} 0 & m_z & m_y \\ m_z & 0 & m_x \\ m_y & m_x & 0 \end{pmatrix} + a_2'\begin{pmatrix} -m_x m_y m_z & -m_z m_x{}^2 & -m_y m_x{}^2 \\ -m_z m_y{}^2 & -m_x m_y m_z & -m_x m_y{}^2 \\ -m_y m_z{}^2 & -m_x m_z{}^2 & -m_x m_y m_z \end{pmatrix}$$

$$\chi^{(2),\mathrm{even}} = a_3'\begin{pmatrix} (m_z{}^2+m_y{}^2) & -m_x m_y & -m_x m_z \\ -m_x m_y & (m_z{}^2+m_x{}^2) & -m_y m_z \\ -m_z m_x & -m_z m_y & (m_y{}^2+m_x{}^2) \end{pmatrix}$$

The decomposition of Edelstein tensors and torkances start from the second order of SOC as follows:

$$\chi_y^y = \chi_y^{y(2)} + \chi_y^{y(3)} + \ldots = \lambda_{yy}^{(2)}(\xi/\xi_0)^2 + \lambda_{yy}^{(3)}(\xi/\xi_0)^3 + \ldots$$

$$\chi_y^z = \chi_y^{z(2)} + \chi_y^{z(3)} + \ldots = \lambda_{zy}^{(2)}(\xi/\xi_0)^2 + \lambda_{zy}^{(3)}(\xi/\xi_0)^3 + \ldots$$

$$t_y^y = t_y^{y(2)} + t_y^{y(3)} + \ldots = \eta_{yy}^{(2)}(\xi/\xi_0)^2 + \eta_{yy}^{(3)}(\xi/\xi_0)^3 + \ldots$$

$$t_y^z = t_y^{z(2)} + t_y^{z(3)} + \ldots = \eta_{xy}^{(2)}(\xi/\xi_0)^2 + \eta_{xy}^{(3)}(\xi/\xi_0)^3 + \ldots$$

Fig.8. (a) and (b) show the dependence of $\chi_y^z$, $\chi_y^y$ and the corresponding $t_y^y$, $t_y^z$ on the expansion of SOC strength by setting $\boldsymbol{m}//x$. $\chi_y^z$ and $t_y^y$ components fit well at second order SOC, while $\chi_y^z$ and $t_y^y$ exhibit non-negligible third order terms. For $\chi_y^y$, the third order coefficient $\lambda_{yy}^{(3)} = 1.932$ is significantly larger than the second order coefficient $\lambda_{yy}^{(2)} = -0.308$, indicating that the third order SOC contribution grows

rapidly with increasing SOC strength. For $t_y^z$, the third order coefficient is comparable with the second order coefficient: $\lambda_{zy}^{(2)} = -0.140$, $\lambda_{zy}^{(3)} = -0.113$.

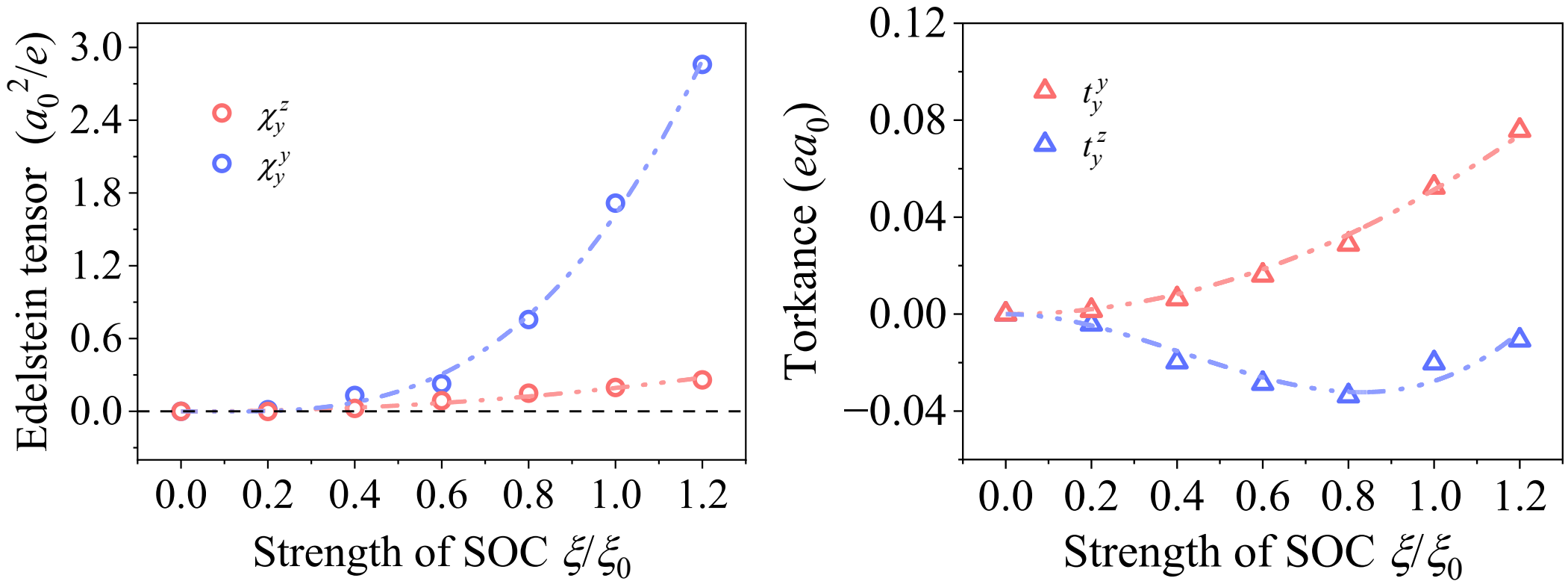

Fig. 8 Edelstein tensors $\chi_y^z$, $\chi_y^y$ and corresponding torkances $t_y^y$, $t_y^z$ as a function of SOC scaling for PtMnSb, with magnetization aligned along the $x$ direction. Open symbols represent the first-principles results, while dashed lines show the corresponding fitting curves. Blue curves: $\chi_y^y$ and $t_y^z$ fitted up to third-order SOC; red curves: fitted with second-order SOC only.

## IV. Summary

In Summary, we develop a spin-group theory of the Edelstein effect and spin-orbit torque in collinear ferromagnets, which enables a systematic expansion of the corresponding response tensors order by order in SOC. For the representative $C_{4v}$ point group, we derive the conventional FL and DL torques, which emerges at both first and second order in SOC. Additional Edelstein tensor components and torkance contributions further arise at second order in SOC. For orbital-Hall-dominated Ti/Ni and spin-Hall-dominated Pt/CoFe bilayers, our first-principles results are fully consistent with the symmetry analysis, demonstrating that irrespective of the underlying microscopic mechanism, spin-orbit torques originate at least at first order in SOC. For the $C_{3v}$ point group, we identify the 3$m$ torque at second order in SOC, which is shown to facilitate deterministic field-free switching of perpendicular magnetization. Finally, we show that in bulk ferromagnetic PtMnSb, a compound with a high-symmetry cubic point group, spin-group symmetry enforces the vanishing of all first-order SOC contributions, such that both the Edelstein effect and spin-orbit torque appear at second and third order in SOC.

Our work establishes a comprehensive spin-group symmetry framework for the Edelstein effect and spin-orbit torque in collinear ferromagnets, and elucidates the leading-order microscopic mechanisms governing these phenomena. These findings advance the fundamental understanding of electrical control of magnetic order, and may pave the way for the design of spintronic devices leveraging higher-order spin-orbit torque effects.

## Acknowledgement

This work was supported by the National Natural Science Foundation of China (Grants No. T2394475, No. T2394470), and the National Key Research and Development Program of China (2024YFA1611200).